\documentstyle[aps,prl,epsf,twocolumn]{revtex}
\begin{document} 
\draft
\title{ Periodic and Aperiodic Bunching in the Addition Spectra of
 Quantum Dots}
\author{N.B. Zhitenev and R.C. Ashoori}
\address{Department of Physics, Massachusetts Institute of 
Technology, Cambridge, 
Massachusetts, 
02139}
\author{ L.N. Pfeiffer and K.W. West}
\address{Bell Labs., Lucent Technologies, Murray Hill, NJ 07974 }
\maketitle
\date{Recieved}
\begin{abstract}
We study electron addition spectra of quantum dots in a broad range
 of electron occupancies starting from 
the first electron. Spectra for dots containing $<$200 electrons
 reveal a surprising feature. Electron additions 
are not evenly spaced in gate voltage. Rather, they group into
 bunches. With increasing electron number 
the bunching evolves from occurring randomly to periodically at
 about every fifth electron. The periodicity 
of the bunching and features in electron tunneling rates suggest
 that the bunching is associated with 
electron additions into spatially distinct regions within the dots.
\end{abstract}

\pacs{PACS 71.30.+h, 73.20.Jc, 73.20Dx, 73.40.Gk}
\narrowtext

Coulomb blockade (CB) is one of the most fundamental and robust concepts in 
 mesoscopic physics. Lambe and Jaklevic first made this point clear
 in a seminal 
experiment nearly 30 years ago\cite{lambe}. For a metallic island poorly
coupled to its surroundings, the number of electrons is quantized at low 
temperatures. 
Due to the repulsive Coulomb potential created by electrons already on the island,
 the energy 
required to add an electron to the island increases by a fixed 
amount $\Delta$ with each 
electron added. An external gate electrode capacitively
 coupled to the 
island through a capacitance $C_g$ can be used to cause electrons to 
transfer on and off the island. Additions 
of  single electrons  occur periodically in gate voltage
 with a period 
$e/C_g$. Physical phenomena in the system with characteristic energy 
scales on order 
$\Delta$ can disrupt the periodicity. For instance superconductivity 
in the system, can 
cause electrons to be added as periodically occurring pairs\cite{tuominen}. 
One does not 
expect such results in a semiconductor or a normal metal sample.

In a semiconductor system containing enough electrons to be 
considered metallic, 
the deviations from  exact periodicity in gate voltage are 
expected to be rather weak.  
The corresponding small parameter is $r_s/R$, where $r_s$ is the screening 
length (about 100~$\AA$), and $R$ is a characteristic size of the system 
(0.2-1~$\mu$m) \cite{houten,sivan,berkovitz,koulakov}. In 
the opposite limit of a disordered mesoscopic system containing a few 
electrons 
localized at spatially distinct sites, significant fluctuations in the 
addition spectrum are 
predicted\cite{koulakov}. 

Several years ago, one experiment on a semiconductor system displayed results 
which appeared to violate CB\cite{ashlg}. Electrons were seen to 
enter a quantum dot in 
pairs rather than individually. The system was a two-dimensional 
dot with a 1~$\mu$m 
diameter. It was somewhat atypical of semiconductor quantum dot 
experiments: the first 
electrons to enter this system occupy random potential minima created
 by disorder, and 
the different minima were screened from each other by a metallic electrode 
which was only 
a few hundred Angstroms away. Nonetheless, the {\it exact} coincidence of 
electron additions 
into the system is quite a surprise. The experiment used a method called
 single electron 
capacitance spectroscopy (SECS) and was unique in allowing study of 
 electron additions 
into separate localized sites.

This letter describes results from a systematic SECS study of different sized 
two-dimensional 
dots to help determine the origin of the strange correlation in electron 
additions. We found that in dots with lithographic diameters larger than 
0.4~$\mu$m containing 
small numbers of electrons, electron additions are sometimes 
grouped in bunches 
comprising from 2 to 6 electrons. The exact pairing seen 
previously\cite{ashlg}  is 
apparently a limiting case of this more general tendency.

As the electron occupancy (N) is increased in dots of all sizes, the bunching 
eventually ceases, and a periodic Coulomb blockade spectrum develops. However, 
application of magnetic field causes the bunching effect to reappear.
 Surprisingly, 
in dots 
with diameters of 0.4-0.5~$\mu$m, the bunching occurs {\it periodically} with 
electron number. Approximately every fifth electron addition peak pairs 
with a neighboring 
peak. The details of the addition 
spectra yield critical clues about the nature of the bunching. 

A schematic of our samples is shown on inset to Fig.1b. They are 
similar to the 
ones described in Refs. \cite{ashlg,ashsm}. The AlGaAs/GaAs wafer 
contains the 
following layers (from the bottom to the top): 3000~$\AA$ $n^+$ GaAs, 
400~$\AA$ 
GaAs spacer layer, 136~$\AA$ AlGaAs/GaAs superlattice tunnel barrier, 
175~$\AA$ 
GaAs quantum well, 500~$\AA$ AlGaAs blocking barrier, 300~$\AA$ GaAs cap
 layer. A 
mesa with deep ohmic contacts down to $n^+$ GaAs is initially defined.
 Then a circular 
Cr/Au gate electrode is fabricated on the top of the mesa.  8 dots were
 studied with
 gate 
diameters ranging from 1.6~$\mu$m to 0.2~$\mu$m. Plasma etching produces 
a short
 pillar (300~$\AA$ tall) 
using the gate electrode as a mask.  Electrons remain in the quantum well
 only in the 
region below the pillar\cite{ashlg}. A larger overlapping metal electrode
 then
 provides electrical connection to the gate.   The measurements are carried
 out using an on-chip bridge circuit described in\cite{ashlg}.

Fig.1a displays the  electron addition spectrum at zero magnetic field for a
 dot of 500~nm lithographic diameter. For 
gate biases below $-500$~mV, the quantum dot is empty. Peaks occur in the
 capacitance at 
gate voltages for single electron additions to the quantum
 dot\cite{ashlg}. Fig. 1b 
displays the Fourier transform (FT) of the measured capacitance
 for successive gate 
voltage intervals of equal  length.  For large N (bottom traces), only
 a single dominant 
frequency component is present in the spectrum. The position of the FT
 peak counts the 
number of electron additions per gate voltage interval. As the gate bias
 is made 
more negative, fewer electrons are added in a gate voltage interval,
 reflecting 
a decrease in the $C_g$ due to lateral contraction 
of the electron droplet. At voltages more negative than $-400$~mV, the single peak
 evolves into a broad 
low frequency spectrum. The broadening  indicates that the gate 
voltage spacings between 
electron additions become uneven.

The peak positions in Fig.1b can be recalculated to determine the 
area of the dot 
using a simple parallel plate capacitor model. This scale is shown 
on the top axis. 
Altogether, we can distinctly resolve about 600 electron additions 
in this dot. The gate 
voltage scale can be directly converted to an energy scale
 $\Delta E=\alpha \Delta V_g$ 
with the lever-arm $\alpha\!\sim$0.5 for these structures
 determined from the geometry of 
the dot as described in\cite{ashlg}. The gate voltage position of the Nth
 capacitance peak, 
when multiplied by the lever arm, directly measures the chemical
 potential $\mu _N$ of the 
dot containing N electrons\cite{houten} .

The magnetic field evolution of a portion of the electron addition spectrum is 
shown in Fig.2a. The grayscale map displays the first 150 additions,
 with capacitance peaks visible as black traces.  
Examination of the bottom of Fig.2a shows that the first 7 electrons 
enter the dot at 
widely spaced voltages. They may enter into a single potential minimum or
 minima 
spaced closely enough that the Coulomb repulsion between the sites is
 sufficient to keep 
the peaks widely spaced. Beyond the 7th electron trace, something
 extraordinary occurs. 
Three electrons enter the dot in very rapid succession in gate voltage
 over the
 full range of 
magnetic fields. The next two electrons also join in a bunch (pair). 
For higher N, other 
bunches can be seen. We note that the experiment shows no hysteretic 
effects. The 
bunching is a phenomenon which occurs with the dot in {\it equilibrium} 
with its 
surroundings.

After about 40 electrons are added to the dot, the bunching develops into a 
periodic pattern, with one bunch appearing for each 4-6 electrons 
added to the dot. As N 
is increased beyond about 80, the bunching ceases for zero magnetic 
field. Instead, the electron additions occur with nearly perfect
 periodicity, as
 is typical of 
CB. However, for nonzero magnetic field strengths, the bunching 
phenomenon returns. 
Bunches again occur periodically in gate voltage, and the period is
 about the same
 as that 
for the zero field bunches. A zoom-in of this behavior is shown in 
Fig. 2b. The onset of 
bunching shifts to larger magnetic fields with increasing concentration,
 and the bunches 
are no longer observable at fields up to 13 Tesla for more than about
 200 electrons
 in the 
dot.

The behavior of each electron trace can be described roughly as follows.
 The 
magnetic field at which all electrons fall into the lowest Landau
 level, $\nu$=2, can be 
readily identified as a maximum in the traces at around
 B=2~T \cite{ashnat}. As in 
two-dimensional systems the chemical potential peaks just as 
higher Landau levels depopulate completely. Jumps in the traces at higher
 magnetic fields, 
where both spin levels of the lowest Landau level are filled, are usually 
interpreted  as 
single electron spin-flips \cite{mceuen,ashnat}. The  flatness of the
 traces around 
B=6~T demarcates total spin polarization of the dot. We refer to this range as the
vicinity of $\nu$=1. For higher fields, 
the traces rise 
nearly linearly with magnetic field.

A bunched pair of traces in Fig.2b is marked with a *. These traces are fairly 
representative of all of the other traces which appear as electron pairs.
 Starting at some nonzero magnetic field the two traces are seen to stick 
together but then they split 
as the field approaches that which yields $\nu$=1. Passing through  $\nu$=1,
 the lower 
trace of the bunched pair splits from the trace above it, only to join with the
 trace below 
it.
 
There is a region devoid of electron additions in the spectrum on both 
sides of any 
bunch. The mean interval between electron additions hence remains the same on
 a larger scale, 
even though individual traces have bunched. This is clearly noticeable 
for small N.
At large N, Fig. 2b reveals that the spacing between the non-bunched 
additions at nonzero 
magnetic field is larger than the spacing at zero field where the 
bunching phenomenon is 
not present.

The bunching appears to be a universal behavior. We have seen the 
bunches in all 
investigated quantum dots with lithographic diameters greater 
than 0.4~$\mu$m. At low 
N, the bunching emerges without any apparent pattern. With 
increasing N, the spacing
 between peaks becomes 
 regular at zero magnetic field, but application of sufficiently 
strong field
 revives the 
bunching. The boundary for the onset of the bunching is remarkably 
similar for all
 dots in 
which bunches are observed, regardless of their size. This boundary 
moves to higher 
magnetic fields as the average electron density  (note, {\it not} N)
  in the dot is 
increased roughly 
according to the linear relation

\begin{equation}
\label{eq1}
n_{onset} =(1.1+0.08\times B[Tesla])\times 10^{11} cm^{-2}
\end{equation}

The nearly periodic bunching 
(pairing) pattern is observed for dots created with lithographic 
diameters of about 
0.5~$\mu$m. For larger dots the bunching still occurs as the magnetic field
 surpasses the 
threshold (\ref{eq1}), but the bunches appear to occur randomly with 
gate voltage rather 
than periodically.

The bunching phenomenon is reflected in the rate at which the electrons tunnel
 into the dot. At zero 
magnetic field, the rate of electron tunneling between the $n^+$ substrate 
and the 
quantum well is about 5~MHz.  Measurements at a much lower frequency 
of f=200~KHz are only sensitive to the tunneling resistance if the 
tunneling is 
strongly suppressed by electrons correlations within the 
dot\cite{ashsm,ashnat}.
 At very low temperatures (T$<$0.1~K) the tunneling rate 
drops substantially in particular regions of magnetic field and electron
 occupancy.

Figure 2c shows a measurement of the addition spectrum of the same
 dot at base 
temperature T=30~mK after thermal cycling up to room temperature. 
The details of the 
addition spectrum of the dot are modified but the overall bunching 
behavior remains 
qualitatively unchanged. For low N, shown on bottom part of Fig. 2c, 
contrast in all 
electron traces is the same over the entire range of magnetic field, 
indicating that the 
electron tunneling rate is much larger than the measurement frequency. 
The middle 
segment of  Fig.2c displays the capacitance spectrum in a range of 
larger N (75-95 
electrons in the dot). Notice here that some of the traces extinguish 
as magnetic field 
increases. As the peaks diminish in strength, the phase of the electron 
tunneling signal lags relative to the ac excitation\cite{ashsm}. This 
detectable
 decay of the 
tunneling rates begins in the vicinity of $\nu$=1, for sufficiently large
 number of electrons 
in the dot. 

The only traces observable at the highest magnetic field of B=13 T in Fig.2c 
extend from paired traces. Examination of  the intensity and phase of these
 unextinguished traces shows that they 
typically result from only a single electron rather than two electrons 
tunneling.
  We note that the dc bias in the experiment is adjusted very slowly so that 
the electron occupancy in the dot changes even though peaks are not seen 
in the 
capacitance experiment. Finally, at higher N (Fig. 2c, upper part), the 
bunching
 disappears, 
and  all traces extinguish equally. 

We believe that pairs of electrons in the quantum dot observed previously by 
Ashoori et al.\cite{ashlg} are a special case of the bunches in the 
regime of electrons 
strongly localized within a large (1~$\mu$m lithographic diameter) dot. In
 dots of similar 
size, we have seen more examples of bunches with the traces of two and 
sometimes 
three electrons that exactly overlap over a range of magnetic fields. 
In general, paired 
 traces from 
dots with smaller lithographic diameters  do not coincide exactly. Two
 theoretical 
models have been suggested to explain the origin of the pairs. 
In one model\cite{wan}, 
lattice deformation mediates the effective attraction in the pairs. In another 
model\cite{raikh}, redistribution of electrons in the dot makes the energy for
 adding two 
electrons less than twice the energy for adding a single electron. Both of these models 
predict a dramatic suppression of the tunneling rate as soon as two electrons are joined 
into a pair, since both electrons must be added into the dot in a coherent
 fashion. Having 
studied a large number of exact pairs in the frequency range 50~KHz - 1~MHz we have 
never observed a significant drop of the tunneling 
rate when the traces merge. This suggests that the paired 
electrons tunnel into 
the dot {\it independently}, though they are added to the 
system at precisely the same gate voltage. Remarkably, the data
 indicate that 
 filling one state of a pair has no effect on 
the energy of the other state in the pair.

A coarse comparison of addition spectra at small and large N leads us to conclude 
that the bunches are intrinsically associated with electron localization
 within the quantum 
dot. The strong fluctuation of the 
spacings between electron additions seen at low N (though not the pairing)
 is expected in 
the strongly localized regime\cite{koulakov}. The development of a periodic CB
 spectrum visible at higher N at zero magnetic field 
indicates the metallic character of the dot in this regime.  The periodic
 bunching occurs in the 
transition range between these regimes. 

The reproducibility of the periodic bunching pattern in several dots cannot be 
ascribed to a peculiarity of the disorder potential. As we observe the same 4-5 electron 
periodicity of the pairing over the entire magnetic field range, we start
 our analysis from a 
consideration of electrons in a dot in infinite magnetic field. Electrons can
 then be considered as classical point charges. In a 
two-dimensional system of uniform potential, classical point charges rest on a
 triangular lattice. 
When the lattice is restricted within a quantum dot, mostly the outer row of
 electrons is 
deformed to conform to the cylindrical symmetry of the confining potential 
\cite{bedanov,levitov}. Electrons in the center of the dot maintain the 
triangular
 lattice 
arrangement. The sequence of the electron entrances into the classical dot
 can be 
calculated. Levitov\cite{levitov} demonstrated recently that for 
 50-150 
electrons in the dot, one electron of each 4-5 electrons added
 enters the outer 
row (circumference). This coincides with the frequency of the 
periodic bunches in our data. Intriguingly, though the classical point charges
 cannot describe the electron density in zero magnetic field, the same 
periodicity of 
electron bunching is experimentally observed down to zero magnetic field.

The idea that one of the bunched electrons appears at the edge of the electron 
droplet is  consistent with the contrast observed in the tunneling rate. 
The drop of the 
tunneling rate for $\nu <$1 illustrated on Fig.2c indicates poor overlap of the
N~-~electron ground state with the (N+1)~-~electron ground state in the dot.
 Such a decay of the tunneling rate can be
 considered 
as a special case of the Coulomb gap observed for larger 
systems\cite{ashgap} . The 
origin of the tunneling suppression can be understood semi-classically. The 
tunneling 
process suddenly adds one more electron into a dot. Until the system
 relaxes to its new ground state, the 
tunneling process is not finished. Hence, the effective tunnel barrier
 depends on the 
disturbance of the density distribution and its relaxation rate. The higher
 tunneling rate 
observed for one trace in each bunch can be explained if that electron is 
introduced into the edge of the dot. That electron then 
has fewer neighbors (and 
they 
are farther away) in comparison with an electron introduced into the bulk of
 the 
dot. 

The mechanism creating the bunching phenomenon remains highly speculative.
  What
 can compete with the usually dominant Coulomb addition energy to disturb 
the addition spectrum so profoundly? Hartree-Fock calculations 
demonstrate\cite{wen} 
that exchange can mediate a local attraction between electrons, tending
 to keep the system 
compact.  The $\nu$=1 state is believed to be 
fully spin-polarized, and exchange maintains $\nu$=1 as the
 lowest energy
 state of the system over a 
range of magnetic fields\cite{wen}. The switching of the bunches at $\nu$=1 
(Fig.2b) appears to effectively broaden this range 
for some of the traces suggesting the involvement of the exchange interaction 
 in bunch formation. 

We gratefully acknowledge numerous useful discussions with Leonid Levitov. 
Expert etching of samples was performed by S.J. Pearton and J.W. Lee. This 
work is supported by the ONR, the Packard Foundation, JSEP-DAAH04-95-1-0038,
 NSF DMR-9357226 and DMR-9311825, and DMR-9421109.

\begin{figure}

\caption{ a) Quantum dot capacitance as a function of gate voltage. 
The peak at -500 
mV denotes the appearance of the first electron in the dot. T=0.3 K. b) 
Fourier transform spectra of 
measured capacitance for  succesive gate voltage intervals of 28 mV starting 
from -500 mV at the 
top to -80~mV at the bottom.
 Peaks in spectra correspond to the number of electrons added 
into the dot per given 
voltage interval. Inset: the scheme of the dot.}

\end{figure}

\begin{figure}

\caption{ Grayscale image of the measured capacitance. Black denotes capacitance
 peaks. Electron occupancies are indicated as numbers. a) Vertical axis - gate 
voltage ranging from -511~mV (bottom) to -328~mV (top). T=0.3 K. b) Zoom~-~in  
 of  spectrum 
surrounded by box in a). c) Segments of the addition spectrum measured after thermally 
cycling the dot to room temperature at T=50 mK. Vertical bar corresponds to energy
 change of 5meV calculated from the gate 
voltage using constant geometrical factor  (common for all images in c) ).}
\end{figure}

\end{document}